\documentclass[a4paper,11pt]{article}
\usepackage{jcappub} % for details on the use of the package, please

       \title{Stochastic gravitational wave background: birth from string-wall death}
	
	\author[a,b]{Shuailiang Ge}
	\affiliation[a]{Center for High Energy Physics, Peking University, Beijing 100871, China}
	\affiliation[b]{School of Physics and State Key Laboratory of Nuclear Physics and Technology, Peking University, Beijing 100871, China}
    \emailAdd{sge@pku.edu.cn}

\abstract{
We study a new source of stochastic gravitational wave background (SGWB) from the final collapse of a network of topological defects. 
Typically, the final collapse is considered negligible for generating gravitational waves (GWs) due to its subdominance compared with the network's long-term evolution in the scaling regime.
However, in some cases, a network can be driven outside of horizon by inflation and later re-enter horizon. Then, the network's final collapse after re-entering horizon becomes the dominant GW source and therefore cannot be neglected. We demonstrate this phenomenon in the context of $N_{\rm DW}=1$ string-wall networks which naturally arise in axion models, although the framework can be generalized to other types of topological networks. The final collapse of walls bounded by strings releases GWs. Our calculation of the corresponding GW spectrum suggests it could be related to the first few bins of the nano-Hertz SGWB signal possibly detected by various Pulsar Timing Array (PTA) collaborations. 
However, it is important to note that such GW spectrum falls within a relatively narrow frequency range, which may not completely account for the PTA signal that spans more than one order of magnitude in frequency.
Furthermore, with different parameter choices, the resultant GWs generated in this mechanism could be probed by various GW interferometry experiments.
}

\begin{document}
\maketitle
\flushbottom

%%Start here

%\noindent \textit{\textbf{Introduction.}}--
\section{Introduction}
% new era LIGO...

% Various ongoing and prospective terrestrial and space ...

% On the other hand, PTA...

% Detecting gravitational waves (GWs) would provide a unique opportunity to probe the early Universe and the related new physics. 

% Evidence of SGWB could be explained by

% ----

% The successful detection of gravitational waves (GWs) marks the beginning of a 

% ----

Detecting gravitational waves (GWs) would provide a unique opportunity to probe the early Universe and the related new physics.
Pulsar timing arrays (PTAs) are powerful in detecting GWs at nano-Hertz (nHz) frequencies. 
Recently, positive evidence of a nHz stochastic GW background (SGWB) with the feature of Hellings-Down correlation is published by several PTA collaborations~\cite{NANOGrav:2023gor, NANOGrav:2023hde, NANOGrav:2023hvm, Antoniadis:2023lym, Antoniadis:2023puu, Antoniadis:2023ott, Reardon:2023gzh, Reardon:2023zen, Zic:2023gta, Xu:2023wog}, more convincing than the previous results~\cite{Arzoumanian:2020vkk, Goncharov:2021oub, Chen:2021rqp, Antoniadis:2022pcn}.
Such an SGWB could be sourced by supermassive black hole binaries~\cite{Vaskonen:2020lbd, DeLuca:2020agl, Kohri:2020qqd, Shen:2023pan}, or by phenomena in the early Universe such as strings~\cite{Blasi:2020mfx, Ellis:2020ena, Samanta:2020cdk, Bian:2022tju, Buchmuller:2023aus, Servant:2023mwt, Lazarides:2022jgr, Antusch:2023zjk}, domain walls~\cite{Bian:2022qbh, Ferreira:2022zzo, Gelmini:2023kvo, Zhang:2023nrs, Du:2023qvj, Babichev:2023pbf}, first-order phase transitions~\cite{Nakai:2020oit, Addazi:2020zcj, Ratzinger:2020koh, Xue:2021gyq, Addazi:2023jvg, Jiang:2023qbm, An:2023jxf, Ahmadvand:2023lpp}, inflationary fluctuations~\cite{Ananda:2006af, Baumann:2007zm, Kohri:2018awv, An:2023idh, HosseiniMansoori:2023mqh, Vagnozzi:2020gtf, Benetti:2021uea, Vagnozzi:2023lwo, Choudhury:2023kam, Oikonomou:2023qfz, Cai:2023dls, Balaji:2023ehk}, other exotic phenomena~\cite{Depta:2023qst, Gouttenoire:2023nzr, Ye:2023xyr, Bousder:2023ida, Huang:2023chx, Yang:2023aak}, etc.

% Detecting gravitational waves (GWs) would provide a unique opportunity to probe the early Universe and the related new physics. 
% Pulsar timing arrays (PTAs) are powerful in detecting GWs at nano-Hertz (nHz) frequencies.
% Recently, several PTA collaborations, including NANOGrav~\cite{NANOGrav:2023gor, NANOGrav:2023hde, NANOGrav:2023hvm}, EPTA~\cite{Antoniadis:2023lym, Antoniadis:2023puu, Antoniadis:2023ott}, PPTA~\cite{Reardon:2023gzh, Reardon:2023zen, Zic:2023gta}, and CPTA~\cite{Xu:2023wog} have published positive evidence of a nHz stochastic GW background (SGWB) with the feature of Hellings-Down correlation, more convincing than the previous results~\cite{Arzoumanian:2020vkk, Goncharov:2021oub, Chen:2021rqp, Antoniadis:2022pcn}. Such an SGWB can be sourced by supermassive black hole binaries~\cite{Vaskonen:2020lbd, DeLuca:2020agl, Kohri:2020qqd}, or by phenomena in the early Universe such as strings~\cite{Blasi:2020mfx, Ellis:2020ena, Samanta:2020cdk, Bian:2022tju}, domain walls~\cite{Bian:2022qbh, Ferreira:2022zzo}, first-order phase transitions~\cite{Nakai:2020oit, Addazi:2020zcj, Ratzinger:2020koh, Xue:2021gyq}, and large curvature perturbations~\cite{Ananda:2006af, Baumann:2007zm, Kohri:2018awv}, etc. 
% See also a series of recent papers~\cite{} attempting to interpret the SGWB mainly within the above frameworks.  

Topological defects, such as strings and domain walls, are widely predicted in various new-physics models with symmetry breakings~\cite{Vilenkin:2000jqa}.  After formation in the early Universe, a string network, a domain wall network or a string-wall network will soon enter the scaling regime~\cite{Albrecht:1984xv, Bennett:1987vf, Allen:1990tv, Press:1989yh, Garagounis:2002kt, 1990ApJ...357..293R, Gorghetto:2018myk}, where the average distance between topological defects is comparable to the Hubble radius. The network dynamics during the scaling regime will emit GWs and form an SGWB~\cite{Blasi:2020mfx, Ellis:2020ena, Samanta:2020cdk, Bian:2022tju, Bian:2022qbh, Ferreira:2022zzo}.

In this work, we propose a new SGWB source from the final disappearance stage of a topological network. 
Compared with the scaling regime which has been extensively studied in the literature~\cite{Hiramatsu:2013qaa, Hiramatsu:2012sc,
Hiramatsu:2012gg, Kawasaki:2014sqa,
Gorghetto:2021fsn, Chang:2019mza, Chang:2021afa}, GWs emission from such a final disappearance stage has been largely overlooked. 
This is understandable since the disappearance process, which marks the death of the network, is relatively short compared with the long-term evolution in the scaling regime. However, as we will see below, in some cases the network's final stage becomes the \textit{only} GW source and thus deserves a careful study. In the following, we call the disappearance process the death or collapse of the network. We will discuss this in the well-motivated axion framework with the domain wall number $N_{\rm DW}=1$~\cite{Vilenkin:1982ks, Sikivie:1982qv}, although the idea can be broadly applied to other types of networks with appropriate adaptations.

Let's first briefly review the traditional axion cosmology.
In QCD axion~\cite{peccei1977cp, peccei1977constraints, weinberg1978new, wilczek1978problem, dine1981simple, Zhitnitsky:1980tq, kim1979weak, shifman1980can} and axion-like particle models~\cite{Ringwald:2014vqa}, axionic strings and walls will naturally emerge in the early Universe~\cite{Vilenkin:1982ks, Sikivie:1982qv}. Such topological defects are usually discussed in two scenarios, depending on whether the $U(1)$ Peccei-Quinn (PQ) symmetry breaks before or after inflation~\cite{Sikivie:2006ni}. In the pre-inflationary scenario, the formed axion strings will be blown away by inflation and no walls will form later since the axion field has also been homogenized. In the post-inflationary scenario, strings form and evolve in the scaling regime until the wall formation when the axion mass effectively turns on. A string is attached to a wall for the case $N_{\rm DW}=1$, and the network quickly collapses under the wall tension. The dominant part of GW emission is from the scaling regime. 

However, in addition to the above pre- and post-inflationary scenarios, a less explored but also very natural and compelling scenario is that PQ symmetry breaks \textit{during} inflation (see e.g., Refs.~\cite{Harigaya:2022pjd, Redi:2022llj}). In this scenario, strings are formed during inflation and will be quickly blown out of inflationary Hubble horizon $H_I$. Since the distances between strings are well beyond $H_I$, there is no string dynamics at all and thus no scaling regime and no GWs. The domain walls form 
% at $m_a H(t) \sim 1$ 
when strings are still super-horizon so they do not collapse. As long as the PQ symmetry breaks not too early during inflation, the formed string-wall network will re-enter horizon, immediately after which the network collapses if the wall energy dominates the network. Therefore, such collapse process becomes the only GW source in this framework. 

In this work, we show that if the re-entering temperature $T_{\rm en}$ is $\mathcal{O}(10)~{\rm MeV}$, the GW emission from the network death could be related to the nHz SGWB observed by PTA collaborations~\cite{NANOGrav:2023gor, NANOGrav:2023hde, NANOGrav:2023hvm, Antoniadis:2023lym, Antoniadis:2023puu, Antoniadis:2023ott, Reardon:2023gzh, Reardon:2023zen, Zic:2023gta, Xu:2023wog, Arzoumanian:2020vkk, Goncharov:2021oub, Chen:2021rqp, Antoniadis:2022pcn}, especially the first few bins of observations. We also demonstrate that for different parameter choices, GWs emitted in this mechanism could be probed by various prospective interferometry experiments at higher frequencies~\cite{Kawamura:2011zz, Crowder:2005nr, LISA:2017pwj, Ruan:2018tsw, TianQin:2015yph, Liang:2021bde, Maggiore:2019uih, Reitze:2019iox, LIGOScientific:2014pky, VIRGO:2014yos, KAGRA:2018plz}. In the following, we first describe the axion model with the feature of re-entering horizon. Then, we calculate the GW spectrum generated by the network collapse. Next, we discuss the cosmological implications of this picture. Finally, we summarize the work and make further discussions.

% Compared with gauge strings which release energy dominantly as GWs, the axionic string-wall network mainly releases energy as free particles in addition to a small portion as GWs. 

%\noindent \textit{\textbf{Model.}}--
\section{Model}
We start with the general Lagrangian that can give birth to the string-wall network with $N_{\rm DW}=1$, 
\begin{equation} 
\begin{aligned}\label{eq:La}
\mathcal{L} \supset -\frac{1}{4}\lambda (\Phi^{\dagger} \Phi - f_a^2)^2 - m_a^2 f_a^2 \left[1-\cos(a/f_a)\right].
\end{aligned}
\end{equation}
The complex scalar field $\Phi = \left|\Phi \right|{\rm e}^{ia/f_a}$ obeys a $U(1)$ global symmetry denoted as the PQ symmetry. The angular variable $a$ is the axion field and $f_a$ is called the axion decay constant. The first term in~\eqref{eq:La} spontaneously breaks $U(1)$ when the Universe temperature drops below $\sim f_a$, which produces strings. The second term explicitly breaks $U(1)$ with the true vacuum $a/f_a=0, 2\pi$, which produces domain walls. In the standard QCD axion models~\cite{peccei1977cp, peccei1977constraints, weinberg1978new, wilczek1978problem, dine1981simple, Zhitnitsky:1980tq, kim1979weak, shifman1980can}, the second term stems from non-perturbative QCD instanton effects, and the relation $m_a f_a\sim m_{\pi}f_{\pi}$ is fixed ($m_{\pi}$: pion mass; $f_{\pi}$: pion decay constant). As we are considering axion-like particles in a general context, we do not restrict ourselves to this relation.
% the $m_a-f_a$ relation does not hold. 
Nonetheless, many axion-like models are well-motivated: some models~\cite{Rubakov:1997vp, Berezhiani:2000gh, Hook:2014cda, Fukuda:2015ana, Dimopoulos:2016lvn, Hook:2019qoh, Kelly:2020dda} prefer heavy axions which can still solve the strong CP problem while simultaneously avoid the axion quality problem~\cite{Kamionkowski:1992mf, Barr:1992qq, Ghigna:1992iv, Holman:1992us}. In addition, axions are also predicted in string theory~\cite{Svrcek:2006yi}.

\begin{figure}
    \centering
    \includegraphics[width=0.8\linewidth]{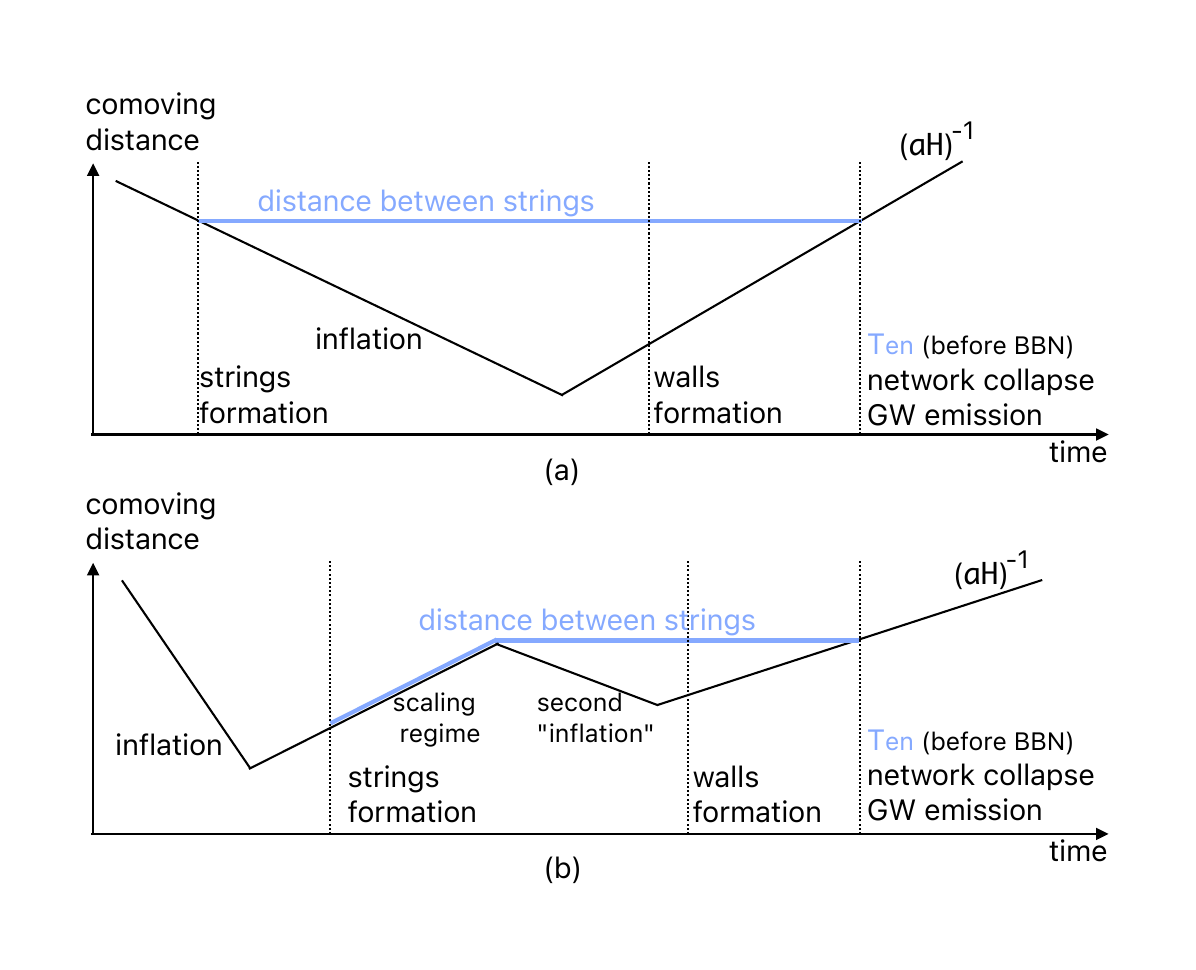}
    \caption{Two possible pictures where the topological network's final collapse (rather than the scaling-regime dynamics) dominates the GW emission. (a) PQ symmetry breaks during inflation. (b) A second inflation follows the PQ symmetry breaking.}
    \label{fig:comoving}
\end{figure}

As discussed in Introduction, we consider the scenario that strings form \textit{during} inflation~\cite{Harigaya:2022pjd, Redi:2022llj} and later the network re-enters horizon so the network collapse becomes the only GW source. If the re-entering temperature is $T_{\rm en}\sim \mathcal{O}(10)~{\rm MeV}$,
the resultant GWs are around the nano-Hertz frequencies.
Supposing that PQ symmetry breaks after $N_{\rm PQ}$ e-foldings of visible inflation, we have~\cite{Redi:2022llj}
\begin{equation} 
\begin{aligned}
T_{\rm en} \simeq T_0 {\rm e}^{N_{\rm PQ}+4}.
\end{aligned}
\end{equation}
where $T_0$ is the current temperature and the factor $4$ is due to the history after matter-radiation equality. Therefore, $T_{\rm en}\sim 10~{\rm MeV}$ corresponds to $N_{\rm PQ}\sim 20$. For a clearer view, we sketch the picture in Fig.~\ref{fig:comoving}(a).

The above during-inflation scenario is natural if the PQ field $\Phi$ couples to either the inflaton field $\phi$ or $H_I$~\cite{Harigaya:2022pjd, Redi:2022llj} (see also Ref.~\cite{An:2023idh}), and thus the PQ symmetry breaking is driven by inflation. For example, one can explicitly build the coupling term $\sim y \phi^2 \Phi^{\dagger}\Phi$ with the coupling $y$, so that the excursion of $\phi$ from large field drives the PQ symmetry breaking when $\phi = \sqrt{\lambda/2y} f_a$. Also, for the case that $H_I$ varies significantly during inflation, one can build a coupling with the Ricci scalar $R$, $-\xi R\Phi^{\dagger}\Phi = 2H_I^2 \Phi^{\dagger}\Phi$, where $\xi=1/6$ if conformal symmetry is imposed and $R = -12H_I^2$ during inflation.

In addition to the above picture, another way to significantly weaken the scaling regime while making the collapse process dominant is to assume a second small inflation after strings have formed, while keeping the first inflation the standard one~\cite{Harigaya:2022pjd}. The second inflation can be a slow-roll inflation or any other kinds of inflations\footnote{
For example, the thermal inflations~\cite{Yamamoto:1985rd, Lyth:1995ka} which are usually associated with the first-order phase transitions. Also, a strict inflation is not necessary as long as the equation of state $w$ is small enough close to $-1$ which can arise in many particle-physics models and beyond (see e.g., Refs.~\cite{Ramberg:2019dgi, Ramberg:2020oct}).   
}.
For a clearer view, we also sketch this picture in Fig.~\ref{fig:comoving}(b).

%\noindent \textit{\textbf{Gravitational-wave spectrum.}}--
\section{Gravitational-wave spectrum}
The string-wall network is made of walls bounded by strings.
% The string-wall network re-enters the horizon at time $t_{\rm en}$. 
The typical size of walls bounded by strings at $T_{\rm en}$ is $R(T_{\rm en})\sim H^{-1}(T_{\rm en})$. Immediately after $T_{\rm en}$, a wall bounded by string quickly collapses under the wall tension $\sigma_{w}$ which dominates the network evolution,
% compared with the string tension $\mu_s$ since the wall carries most of the energy, 
$\sigma_{w} H^{-1}(t_{\rm en})\gg  \mu_s$~\cite{Chang:1998tb}, where $\mu_s$ is the string tension. The collapse speed will be relativistic. 
% The collapse of walls bounded by strings will emit gravitational waves. 
One can conveniently conceptualize it as a shrinking thin massive disk. The energy of a wall bounded by string varies during collapse, $M\sim R^2(t) \sigma_{w}$, predominantly released in the form of free axions, and a small fraction of energy is released in the form of GWs~\cite{Hiramatsu:2013qaa, Hiramatsu:2012sc, Hiramatsu:2012gg, Kawasaki:2014sqa}. 

Based on the simulations of the evolution of $N_{\rm DW}=1$ axionic string-wall network~\cite{Chang:1998tb, Hiramatsu:2012gg, Kawasaki:2014sqa}, we can model the evolution of a wall bounded by string as
\begin{equation} 
\begin{aligned} \label{eq:Rt}
R(t) \simeq R_0 {\rm e}^{-c_R \cdot \frac{\omega_R}{\pi} (t-t_{\rm en})} \cos[\omega_R (t-t_{\rm en})].
\end{aligned}
\end{equation}
$R_0=R(T_{\rm en})$ is the initial size of the object. 
Eq.~\eqref{eq:Rt} is based on Figure 5 in Ref.~\cite{Chang:1998tb}.
It should be pointed out that Ref.~\cite{Chang:1998tb} conducted a 2D simulation, and the distance between ``string and anti-string cores" in Ref.~\cite{Chang:1998tb} corresponds to the size of walls bounded by strings in 3D.
Although Eq.~\eqref{eq:Rt} represents a simplified model, it effectively captures the main dynamics of the walls bounded by strings.
The cosine function represents the oscillating feature;
% of the wall bounded by string; 
the exponential term represents the dominant energy loss into free axions. 
The collapse of walls bounded by strings in the $N_{\rm DW} = 1$ case has been observed in simulations such as those in Refs.~\cite{Chang:1998tb, Hiramatsu:2012gg, Kawasaki:2014sqa}.
However, I found it most straightforward to extract the dynamics from Figure 5 of Ref.~\cite{Chang:1998tb} for further analytical calculations of corresponding gravitational waves.
The radiated axions have an averaged momentum characterized by the Lorentz factor $\gamma_a = \left<\omega_a\right>/m_a \approx 3.2$~\cite{Hiramatsu:2012gg, Kawasaki:2014sqa}, from which the averaged axion speed $\left<v_a\right>$ can be inferred. The same references also found the trend that $\gamma_a$ is insensitive to the axion parameters for $m_a \ll f_a$. Axions are radiated due to the dynamics of the string boundary unzipping the wall, and the collapse speed can be estimated to be close to the axion speed, so that $\omega_R \sim \pi/2 \cdot \left<v_a\right>/R_0$. Such an estimate is also compatible with the numerical results in Ref.~\cite{Dunsky:2021tih}\footnote{Authors of
Ref.~\cite{Dunsky:2021tih} numerically solved the gauged case (see Figure 12 therein) that walls bounded by strings do not lose energy into free particles, so the evolution has no exponential decay term. But we can still make a comparison because our $c_R/\pi \sim 0.1$ in~\eqref{eq:Rt} is small. Our estimate of wall collapse speed $\left<v_R^2\right> \sim \left<v_a^2\right> \approx 0.9$ is larger than their result $\left<v_R^2\right>\sim 0.6$. This is expected because the energy during collapse in their case is conserved (ignoring the very slow loss into GWs), which buffers the collapse. 
In this sense, we consider the results in both cases to be compatible, with no contradictions between them.
}.

% In addition, Ref.~\cite{Chang:1998tb} shows more details of how the size $R(t)$ evolves. 
It is found that the string cores meet head-on and can go through each other~\cite{Chang:1998tb}.
A negative $R(t)$ in~\eqref{eq:Rt} naturally represents such behavior. Based on Figure 5 in Ref.~\cite{Chang:1998tb}, we observe that the oscillation amplitude decreases to about $2/3$ of the previous oscillation\footnote{
As an analogy, the three-dimensional axionic domain wall bubble loses $\sim 40\%$ energy after each oscillation~\cite{Widrow:1989vj}.
}, 
from which we infer that the damping parameter is roughly $c_R\sim 0.4$. However, it is difficult to obtain the exact value of $c_R$ without more detailed simulations, which is sensitive to $m_a$ and $f_a$. In general, we treat $c_R$ as an $\mathcal{O}(0.1)$ number. Our model of the size evolution~\eqref{eq:Rt} of a wall bounded by string matches well with the result in~\cite{Chang:1998tb} (see e.g., Figure 5 therein).
Further numerical simulations, focusing on the detailed evolution of $R(t)$ during collapse and the corresponding GW emission beyond~Refs.~\cite{Chang:1998tb, Hiramatsu:2012gg, Kawasaki:2014sqa}, would be valuable for getting more exact results. Nevertheless, Eq.~\eqref{eq:Rt} incorporates crucial features of the evolution of a wall bounded by string. 

% Further numerical simulations focusing on the detailed evolution of $R(t)$ during the collapse process and the corresponding GW emission, which are beyond~Refs.~\cite{Chang:1998tb, Hiramatsu:2012gg, Kawasaki:2014sqa}, are necessary for getting more exact results. Nevertheless, our model~\eqref{eq:Rt} incorporates main features of the evolution of a wall bounded by string. For now, it enables us to carry out order-of-magnitude calculations of the emitted GWs.

We can estimate the power of emitted GWs as 
\begin{equation} 
\begin{aligned}\label{eq:Power_GW}
P_{\rm GW}(t) \sim G \dddot{Q}_{ij}  \dddot{Q}_{ij}
,~
Q\sim \mathcal{A} M(t)R^2(t) \sim \mathcal{A} \sigma_{w} R^4(t) 
\end{aligned}
\end{equation}
where $Q$ is the quadrupole momentum of a wall bounded by string. $\mathcal{A}$ is the wall area parameter representing the number of walls per Hubble volume at $T_{\rm en}$, which is expected to be close to $1$. 
Once the walls bounded by strings re-enter horizon and start to oscillate, we regard them as independent objects which have detached from each other. This implies that the number density of walls bounded by strings after re-entering horizon decreases following the relation $\propto a^{-3}(t)$ where $a(t)$ is the scale factor.
Then, the energy density of GWs per unit of time is
% \beq
% d\rho_{\rm GW} \sim \frac{P_{\rm GW}(t) dt}{H^{-3}(t_{\rm en})}\frac{a^3(t_{\rm en})}{a^3(t)} 
% \sim  P_{\rm GW}(t) t_{\rm en}^{-3/2} t^{-3/2} dt.
% \eeq
\begin{equation} 
\begin{aligned}\label{eq:drho_over_dt}
\frac{d\rho_{\rm GW}(t)}{dt} \sim \frac{P_{\rm GW}(t)}{H^{-3}(t_{\rm en})}\frac{a^3(t_{\rm en})}{a^3(t)} 
\sim  P_{\rm GW}(t) t_{\rm en}^{-3/2} t^{-3/2}.
\end{aligned}
\end{equation}
The factor $a^3(t_{\rm en})/a^3(t)$ represents the dilution of walls bounded by strings after re-entering horizon.
In the last step, we have assumed that the collapse happens in the radiation-dominated era.
% after the network re-enters the horizon. 
The frequency of emitted GWs concentrates at $f_{\rm GW}\sim \omega_R/\pi$, which then experiences different redshifts to form a frequency range today,  $f(t_0)=f_{\rm GW} a(t)/a(t_0)$. 
% Although the collapse of walls bounded by strings generates almost monochromatic gravitational waves, 
%they experience different redshifts to form a spectrum. 
Then, we can get 
the present-day spectrum of GWs:
\begin{equation} 
\begin{aligned}\label{eq:Omega0}
\Omega_{\rm GW}(t_0) 
& \equiv \frac{1}{\rho_{\rm cr}(t_0)} \frac{d\rho_{\rm GW}(t_0)}{d\ln{f(t_0)}}
\simeq  \frac{1}{\rho_{\rm cr}(t_0)}  \frac{a^4(t)}{a^4(t_0)}    \frac{d\rho_{\rm GW}(t)}{d t} \frac{1}{H(t)}.   \\ 
\end{aligned}
\end{equation}
$\rho_{\rm cr}$ is the critical energy density.

\begin{figure}
    \centering
    \includegraphics[width=0.8\linewidth]{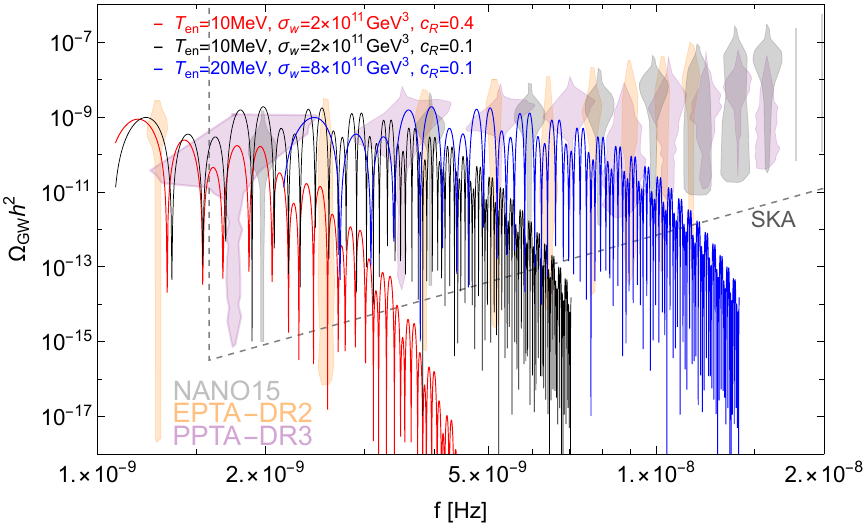}
    \caption{GW spectra $\Omega_{\rm GW}(f)h^2$ from the collapse of walls bounded by strings compared with PTA observations. We choose different values of $T_{\rm en}$, $\sigma_w$, and $c_R$ to show the spectra.
    % : $(T_{\rm en}, \sigma_w, c_R) = 
    % (10~{\rm MeV}, 2\times 10^{11}~\GeV^3, 0.4)$ (red); 
    % $(10~{\rm MeV}, 2\times 10^{11}~\GeV^3, 0.1)$ (black);
    % $(20~{\rm MeV}, 8\times 10^{11}~\GeV^3, 0.1)$ (blue). 
    PTA observations are shown as shaded regions with different colors: NANOGrav-15yr~\cite{NANOGrav:2023gor, NANOGrav:2023hvm}, EPTA full data-release 2 (DR2)~\cite{Antoniadis:2023ott}, and PPTA full data-release 3 (DR3)~\cite{Reardon:2023gzh}, all with Hellings-Down correlations. They are adapted from the corresponding references and converted from the original spectral density to $\Omega_{\rm GW}(f)h^2$ if needed. In addition, the dashed line shows the projected sensitivity of SKA~\cite{Janssen:2014dka}.
    }
    \label{fig:constraints}
\end{figure}

\begin{figure}
    \centering
    \includegraphics[width=0.8\linewidth]{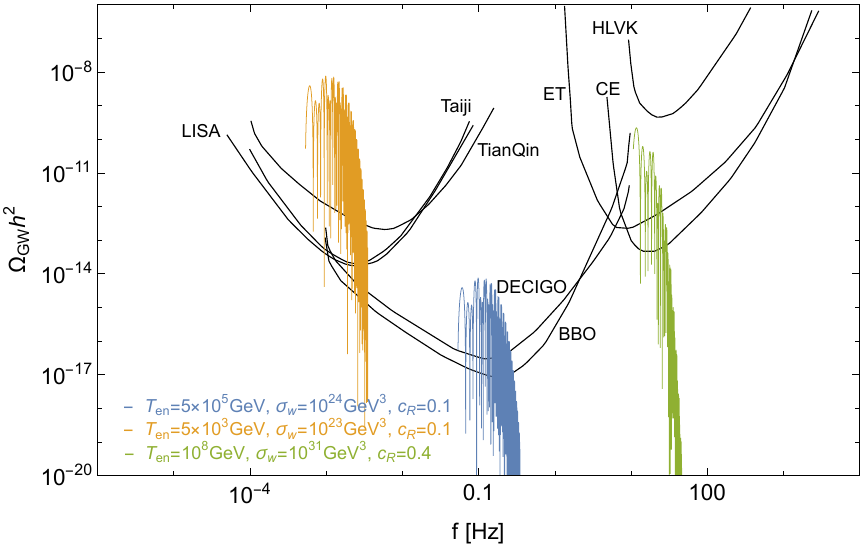}
    \caption{Same as Fig.~\ref{fig:constraints} but for different choices of parameters to show the possibility of being detected by various ongoing and prospective GW experiments with interferometry: DECIGO~\cite{Kawamura:2011zz}, BBO~\cite{Crowder:2005nr}, LISA~\cite{LISA:2017pwj}, Taiji~\cite{Ruan:2018tsw}, TianQin~\cite{TianQin:2015yph, Liang:2021bde}, ET~\cite{Maggiore:2019uih}, CE~\cite{Reitze:2019iox}, and HLVK (including aLIGO~\cite{LIGOScientific:2014pky}, aVirgo~\cite{VIRGO:2014yos}, and KAGRA~\cite{KAGRA:2018plz}). }
    \label{fig:constraints-NEW}
\end{figure}

Combining eqs.~\eqref{eq:Rt}-\eqref{eq:Omega0}, we see that the spectrum only depends on the wall tension $\sigma_w$, the temperature $T_{\rm en}$, and also our choice of the damping parameter $c_R$. In Fig.~\ref{fig:constraints}, we show the spectrum $\Omega_{\rm GW}(t_0)h^2$ for different values of $\sigma_w$, $T_{\rm en}$ and $c_R$. We choose the parameter space inspired by the recent observations on the potential nHz SGWB by various PTA collaborations.
% , NANOGrav~\cite{NANOGrav:2023gor, NANOGrav:2023hde, NANOGrav:2023hvm}, EPTA~\cite{Antoniadis:2023lym, Antoniadis:2023puu, Antoniadis:2023ott}, PPTA~\cite{Reardon:2023gzh, Reardon:2023zen, Zic:2023gta}, and CPTA~\cite{Xu:2023wog}. 
$T_{\rm en}$ determines the GW frequency, and $T_{\rm en} \sim 10-20$~MeV corresponds to nHz GWs. $\sigma_w$ controls the GW-spectrum amplitude, and $\sigma_w\gtrsim 10^{11}~{\rm GeV}^3$ match the PTA observations. $c_R$ controls the span of frequency range. Due to our uncertainty on $c_R$, we chose different values, $0.1$ and $0.4$, based on the discussions above. The collapse of walls bounded by strings is very rapid after $T_{\rm en}$, so the GW frequency range is relatively narrow. This mechanism can potentially explain the first several frequency bins, the most important bins for determining an SGWB, of the current PTA observations.

Comparing with the cases of using scaling-regime networks to explain the nHz SGWB~\cite{Bian:2022qbh, Ferreira:2022zzo}, $\sigma$ here is relatively small. This is due to the fast-oscillation feature which implies a more rapid and efficient GW emission. If considering there is a distribution of the initial size of walls bounded by strings, we expect the oscillating GW spectra will become somewhat flatter, with the averaged amplitude at about $10^{-10}$. 

In addition, by choosing different parameter space of $\sigma_w$, $T_{\rm en}$, and $c_R$, the GW spectra can be potentially detected by various ongoing and prospective GW interferometry experiments in a vast frequency range $\sim 10^{-5}-10^3$~Hz, as shown in Fig.~\ref{fig:constraints-NEW}.

PTA collaborations~\cite{NANOGrav:2023gor, Antoniadis:2023ott, Reardon:2023gzh} have derived posterior probability distributions for the amplitude and slope of GW spectra. 
The frequency range of these spectra can span over more than one order of magnitude, although only the lowest frequency bins carry significant statistical weight. 
In comparison, the GW spectra generated by our mechanism are relatively narrow, because the collapse of walls bounded by strings is a fast process after re-entering horizon. It is possible for such spectra to be related to the lowest frequency bins (the most important ones) in PTA observations, although it seems challenging for them to fully explain the observed spectra. Nonetheless, besides PTAs, the stochastic GW spectra generated by this mechanism could be tested by various GW experiments (e.g., LISA) as shown in Figure~\ref{fig:constraints-NEW}.
Additionally, we want to point out that the GW spectra in this mechanism are influenced by other factors, such as the shapes and size distribution of walls bounded by strings. 
For simplicity, we have treated the shapes of walls bounded by string as flat disks. 
In reality, the shapes could deviate from being flat.
Moreover, the walls bounded by strings could have a distribution, meaning the re-entering horizon times also have a distribution. 
Considering all these effects is expected to make the GW spectrum somewhat flatter and extend the frequencies over a larger range.
A more precise GW spectrum can be obtained with the help of lattice simulations. 
However, simulating the collapse of an axionic string-wall network presents a notorious difficulty due to the existence of two drastically different scales, $m_a\ll f_a$ (or, in another way, the QCD scale $\Lambda_{\rm QCD} \ll f_a$ for the QCD axion case), simultaneously exist in the system. 
This multi-scale problem makes it impossible to perform simulations with realistic values of $m_a$ and $f_a$.
Instead, one has to artificially bring $m_a/f_a$ closer to 1 and then extrapolate the simulation results to the case $m_a/f_a \ll 1$, as done in Refs.~\cite{Chang:1998tb, Hiramatsu:2012gg, Kawasaki:2014sqa}.

%\noindent \textit{\textbf{Comological implications.}}-- 
\section{Comological implications}

Only a small portion of energy stored in the walls bounded by strings is released as GWs. This can be easily seen as follows. The energy carried by GWs $\rho_{\rm GW}(t_{\rm en}) \sim P_{\rm GW} H^{2}(t_{\rm en}) \sim G\sigma_w^2$ is much smaller than the total energy stored in the network $\rho_w(t_{\rm en}) \simeq \sigma_w H(t_{\rm en})$ for the values of $\sigma_w$ chosen above. Most of the energy is released in the form of axions, whose energy density is $\rho_a (t_{\rm en}) \simeq \rho_w (t_{\rm en})$. Due to the small Lorentz factor $\gamma_a$, the radiated axions will soon become cold matter with the Universe's expansion. The proportion of radiated axions accounting for the total energy density is
\begin{equation} 
\begin{aligned}\label{eq:rho-a_over_rho-cr}
\frac{\rho_a (t_{\rm en})}{\rho_{\rm cr}(t_{\rm en})} \sim 10^{-4} \times \left(\frac{\sigma_w}{10^{11}~{\rm GeV}^3}\right)
\left(\frac{10~{\rm MeV}}{T_{\rm en}}\right)^2.
\end{aligned}
\end{equation}
As the Universe evolves, Eq.~\eqref{eq:rho-a_over_rho-cr} implies that the cold axions would overproduce dark matter. This is a common issue shared generally by domain-wall models aiming to explain the nHz SGWB observed by PTAs~\cite{Bian:2022qbh, Ferreira:2022zzo}. Generating the desired PTA SGWB amplitude by walls is accompanied by the overproduction of wall particles. To deal with this issue, the cold particles are usually assumed to further decay into Standard Model (SM) particles or dark radiation~\cite{Bian:2022qbh, Ferreira:2022zzo}. We discuss these two cases separately in the following.

1. Decaying into SM particles.
In our case, it is natural to deplete the axion abundance by decaying into two SM photons (or gluons) via the coupling 
\begin{equation} 
\begin{aligned}\label{eq:aSM}
\mathcal{L}_{a\gamma\gamma {\rm ~or~ } agg}  = \frac{1}{4} \frac{\beta_{\gamma}}{f_a} a F_{\mu\nu} \tilde{F}^{\mu\nu},
~ {\rm or }~
\frac{1}{4} \frac{\beta_g}{f_a} a G_{\mu\nu} \tilde{G}^{\mu\nu}.
\end{aligned}
\end{equation}
$F_{\mu\nu}$ ($G_{\mu\nu}$) is the photon (gluon) field strength and the tilde field is the dual. The coupling strengths are suppressed by the scale $f_a$; $\beta_{\gamma,g}$ is normally an $\mathcal{O}(1)$ number (see e.g., Ref.~\cite{Kelly:2020dda}). The corresponding decay rate is 
\begin{equation} 
\begin{aligned}\label{eq:Gamma}
\Gamma_{a\rightarrow \gamma\gamma, gg} = \frac{\beta_{\gamma,g}^2}{64\pi}\frac{m_a^3}{f_a^2}.
\end{aligned}
\end{equation}
$\Gamma$ should be large enough to deplete the abundance of free axions in time. To avoid potential conflicts with BBN, we conservatively require the axions to decay into the standard model contents before BBN. To ensure this, we adopt the lowest reheating temperature $T_{\rm rh}\simeq 4$~MeV that have been obtained in the literature (see e.g., Refs.~\cite{deSalas:2015glj, Kawasaki:2000en, Bai:2021ibt}) consistent with cosmological observations. The reheating phase in our case corresponds to the period of axion decay. Based on the definition of the reheating temperature, successful BBN processes require that 
\begin{equation}\label{eq:cri_SM_reheating}
    \Gamma_{a\rightarrow \gamma\gamma, gg} \gtrsim 3 H(T_{\rm rh}).
\end{equation}
Plugging the expression \eqref{eq:Gamma} into \eqref{eq:cri_SM_reheating}, we get the requirement for $m_a$ and $f_a$,
\begin{equation} 
\begin{aligned}\label{eq:ma-fa-constraint}
\left(\frac{m_a}{1~{\rm MeV}}\right)^3 \left(\frac{10^6~{\rm GeV}}{f_a}\right)^2 \gtrsim 1.
\end{aligned}
\end{equation}

Beyond this simple argument based on the reheating temperature here, a more detailed calculation of the limits on the abundance of non-relativistic matter during BBN is provided in Ref.~\cite{Yeh:2024ors} (see also the earlier work Ref.~\cite{Scherrer:1987rr}). The paper provided the constraints on the initial abundance and decay rate of a matter-like species $X$ using observational data. In our case, initially the axion abundance at $T_{\rm en}\sim 10$~MeV is about $10^{-4}$ (refer to eq.~\eqref{eq:rho-a_over_rho-cr}). Then, if we apply here the constraints from Ref.~\cite{Yeh:2024ors} (specifically, refer to Figure 8 therein%%%
\footnote{The y-axis of that figure denotes the abundance of the X particle at $10$~MeV as defined in eqs.~(2.5)-(2.7) in Ref.~\cite{Yeh:2024ors}.
}), 
the lifetime of such a particle cannot exceed $\sim 400$ sec. Therefore, $\Gamma \gtrsim 1/(400{\rm~\sec}) $ yields
\begin{equation} 
\begin{aligned}\label{eq:ma-fa-constraint_2}
\left(\frac{m_a}{1~{\rm MeV}}\right)^3 \left(\frac{10^8~{\rm GeV}}{f_a}\right)^2 \gtrsim 1.
\end{aligned}
\end{equation}

Both eq.~\eqref{eq:ma-fa-constraint} and eq.~\eqref{eq:ma-fa-constraint_2} suggest heavy axions, which are not traditional sub-eV QCD axions~\cite{kim1979weak, shifman1980can, dine1981simple, Zhitnitsky:1980tq} with the relation $m_a$-$f_a$ fixed. 
However, heavy axion models are still attractive and well-motivated. For example, heavy versions of QCD axion have been built in a series of models~\cite{Rubakov:1997vp, Berezhiani:2000gh, Hook:2014cda, Fukuda:2015ana, Dimopoulos:2016lvn, Hook:2019qoh, Kelly:2020dda} which can still solve the strong CP problem while simultaneously avoid the axion quality problem. Another example is that heavy axions can arise in the string theory~\cite{Svrcek:2006yi}. 
Furthermore, heavy axions considered here can be searched for at beam dumps, fixed targets, and collider experiments.
For example, the wall tension $\sigma_w=8 f_a^2 m_a\sim 10^{12}~{\rm GeV}^3$ (for PTA observations) together with the constraint~\eqref{eq:ma-fa-constraint} or~\eqref{eq:ma-fa-constraint_2} can be satisfied by the parameter space $m_a \sim (10^{-2}, 10^2)~{\rm GeV}$ with $f_a \sim (10^7, 10^5)~{\rm GeV}$, which can be probed by DUNE ND~\cite{Kelly:2020dda}, HL-LHC~\cite{Hook:2019qoh}, FASER~\cite{FASER:2018eoc}, NA62~\cite{Ertas:2020xcc}, etc. In addition, assuming $m_a$ is temperature-dependent (see e.g., Refs.~\cite{OHare:2021zrq, Arias:2012az, Nakagawa:2022wwm}), similar to the QCD axion cases,  could open up more parameter space.

2. Decaying into dark photons.
Instead of coupling with SM via~\eqref{eq:aSM}, we can assume that axions couple to the dark sector, e.g., dark photon (massless or light enough to behave as radiation),
\begin{equation} 
\begin{aligned}
\label{eq:aDP}
\mathcal{L}_{a \gamma'\gamma'} = \frac{1}{4}\frac{\beta_{\gamma'}}{f_a}aF'_{\mu\nu}\tilde{F}'^{\mu\nu}, 
\end{aligned}
\end{equation}
where the prime field represents dark photon. Such a coupling has been widely used in generating dark photon dark matter by tachyonic instability from axion oscillation~\cite{Agrawal:2018vin, Bastero-Gil:2018uel, Dror:2018pdh, Co:2018lka}. Unlike those references, here we consider the axion decaying into dark photons.

The constraint derived in the previous case based on the reheating temperature, eq.~\eqref{eq:ma-fa-constraint}, can also be applied here. Since the decay product here is dark radiation, we should further check that this does not violate the observational constraints on the extra (radiation) energy. The BBN constraints on the extra radiation are $\rho_{\rm extra}/\rho_{\nu} < 0.3$ below $T_{\rm BBN}\sim {\rm MeV}$ where $\rho_{\nu}$ is the energy density of a single flavor of SM neutrinos~\cite{Planck:2018vyg}. At the reheating temperature $T_{\rm rh} \simeq 4$~MeV, the relative abundance of dark radiation can be estimated as
$\rho_a(T_{\rm rh})/\rho_{\rm cr}(T_{\rm rh})\simeq \rho_a(T_{\rm en})/\rho_{\rm cr}(T_{\rm en}) \cdot T_{\rm en}/T_{\rm rh} \simeq 10^{-3}$ which is far smaller than $0.3$, implying it is safely within the BBN constraints on the extra radiation energy. 

Furthermore, in addition to the case of a matter-like species decaying into standard model contents, Ref.~\cite{Yeh:2024ors} also studied the case of decays into dark radiation. Applying the result from Ref.~\cite{Yeh:2024ors} (specifically, refer to Figure 12 therein), for our case where the initial axion abundance is $\sim 10^{-4}$, the axion lifetime should not exceed $\sim 10^3$~sec which is slightly longer than in the previous case. Therefore, we can obtain a similar constraint on $m_a$ and $f_a$ as in eq.~\eqref{eq:ma-fa-constraint_2}.

One comment on the coupling $\beta_{\gamma'}$ follows.
$\beta_{\gamma'}$ is $\mathcal{O}(1)$ but can be $\mathcal{O}(10-100)$ or even larger via careful model buildings~\cite{Agrawal:2017eqm, Agrawal:2018vin}, so the condition like eqs.~\eqref{eq:ma-fa-constraint} and~\eqref{eq:ma-fa-constraint_2} can be relaxed. 
Compared with the previous case, the coupling between axion and dark radiation, eq.~\eqref{eq:aDP}, is more difficult for detection when assuming that the sector of axion and dark photon decouples from SM.

One bonus is that the resultant dark photons can help alleviate the Hubble tension~\cite{Schoneberg:2021qvd, Riess:2019cxk, riess2021cosmic} since dark photons contribute to the extra energy density of relativistic species.
%effective number of degrees of freedom, $N_{\rm eff}$, of relativistic species. 
As pointed out in Ref.~\cite{Bian:2022qbh}, an interesting coincidence has been observed that in the domain wall models, the same parameter space can simultaneously explain the PTA nHz SGWB and alleviate the Hubble tension significantly. In our case, $\sigma_w \gtrsim 10^{11}~{\rm GeV}^3$ is smaller than that in Ref.~\cite{Bian:2022qbh} as we discussed in the previous section, so the contribution to the extra radiation energy is less. However, the coincidence holds in Ref.~\cite{Bian:2022qbh} under the assumption that free axions (or other scalar particles) 
decay immediately after being released from the network, i.e., $\Gamma_{a\rightarrow \gamma'\gamma'}H^{-1}(T_{\rm en})\gg 1$. In our case, if we assume the decay is not so fast (but still satisfying BBN constraints), the resultant dark photons may still contribute significantly to the extra radiation energy to alleviate the Hubble tension, since the energy in the form of cold axions dilutes slower than in the form of dark photons. There could also be a possibility that the remaining cold axions serve as dark matter. A detailed study of the cosmic evolution of the multiple components is required to answer these questions. We leave this for future work.

Finally, we comment on the formation of primordial black holes (PBHs) when the network re-enters horizon. As pointed out in Ref.~\cite{Ge:2023rrq}, a small portion of the network is closed domain walls, which can collapse into PBHs at $T_{\rm en}$. For the $\sigma_w$ we have chosen, the resultant PBHs could overproduce dark matter. However, this can be avoided if we focus on the mechanism in Fig.~\ref{fig:comoving}(b) where a scaling evolution of strings before the second inflation can significantly destroy the initial conditions of forming closed walls.

\section{Conclusion and discussion}

We have studied the GW emission from the final collapse of domain wall networks, which has long been overlooked compared with the emission during the scaling regime. In the cases with the feature of the network re-entering horizon, the scaling regime vanishes or becomes unimportant, so the network's final collapse would be the dominant GW source. If the re-entering temperature is around dozens of MeV, the resultant GW emission could potentially be related to the first few bins of the nHz SGWB from PTA observations~\cite{NANOGrav:2023gor, NANOGrav:2023hde, NANOGrav:2023hvm, Antoniadis:2023lym, Antoniadis:2023puu, Antoniadis:2023ott, Reardon:2023gzh, Reardon:2023zen, Zic:2023gta, Xu:2023wog, Arzoumanian:2020vkk, Goncharov:2021oub, Chen:2021rqp, Antoniadis:2022pcn}.
A caveat, however, is that this GW emission, which falls within a relatively narrow frequency range, may not fully explain the broader PTA GW spectrum. 
Furthermore, with different parameter choices, the resultant GWs could be detected by various interferometry-based GW experiments~\cite{Kawamura:2011zz, Crowder:2005nr, LISA:2017pwj, Ruan:2018tsw, TianQin:2015yph, Liang:2021bde, Maggiore:2019uih, Reitze:2019iox, LIGOScientific:2014pky, VIRGO:2014yos, KAGRA:2018plz}.

To avoid free particles from the network being overproduced, they are assumed to further decay into SM or dark radiation. 
The constraints require the particle mass to be MeV scale or heavier. Such heavy particles (axions) are theoretically motivated~\cite{Rubakov:1997vp, Berezhiani:2000gh, Hook:2014cda, Fukuda:2015ana, Dimopoulos:2016lvn, Hook:2019qoh, Kelly:2020dda} and could be probed in various experiments~\cite{Kelly:2020dda, Hook:2019qoh, FASER:2018eoc, Ertas:2020xcc}. The case of decaying into dark radiation would help alleviate Hubble tension, which may deserve a careful study by solving the cosmic evolution of multiple components in detail. 

% The GW spectra presented in this Letter derive from order-of-magnitude calculations. 
% They capture the key aspects of a string-wall network's final collapse, yet further detailed numerical simulations are desired for more exact results. 
Finally, we comment that although this work is based on the $N_{\rm DW}= 1$ axion string-wall network, the concept of re-entering horizon can be generally applied to other types of networks~\cite{Blasi:2020mfx, Ellis:2020ena, Samanta:2020cdk, Bian:2022qbh, Bian:2022tju, Ferreira:2022zzo}, which could potentially modify the existing or open up new parameter spaces for generating PTA SGWB, or more broadly, SGWB at different frequencies.

%and use the fixed relation $\sigma_w = 8f_a^2 m_a$

\acknowledgments
	
% \section{Acknowledgments}
I would like to thank Haipeng An, Bin Guo, Jing Shu, and Junchao Zong for discussions. This work is supported by National Natural Science Foundation of China under Grant No. 12247147, the International Postdoctoral Exchange Fellowship Program, and the Boya Postdoctoral Fellowship of Peking University.
	
% \begin{acknowledgments}
% 		%\noindent \textit{\textbf{Acknowledgment.}}--
% \end{acknowledgments}

% \appendix

% \section{XXX} 

%%\clearpage

%     \bibliographystyle{JHEP}
% \bibliography{references.bib}

\providecommand{\href}[2]{#2}\begingroup\raggedright\endgroup

\end{document}